\documentclass[preprint]{revtex4-1}
\usepackage{amssymb}

\usepackage{dcolumn}
\usepackage{amsmath}
\usepackage{color}
\usepackage{ulem}
\usepackage{graphicx}

\usepackage{bm}
\usepackage{amssymb}
\usepackage{soul}
\usepackage{color}
\usepackage{siunitx}
\usepackage{gensymb}
\usepackage{textcomp}
\usepackage{epstopdf}

\def\be{\begin{equation}}
\def\ee{\end{equation}}
\def\bea{\begin{eqnarray}} 
\def\eea{\end{eqnarray}}

\begin{document}

\title{Detection of charge density wave phase transitions at 1T-TaS$_2$/GaAs interfaces}
\author{Xiaochen Zhu, Ang J. Li, G. R. Stewart and Arthur F. Hebard}
\affiliation{Department of Physics, University of Florida, Gainesville, FL 32611, USA}
\begin{abstract}
The transition metal dichalcogenide 1T-TaS$_2$ is well known to harbor a rich variety of charge density wave (CDW) distortions which are correlated with underlying lattice atom modulations. The long range CDW phases extend throughout the whole crystal and terminate with charge displacements at the crystal boundaries. Here we report on the transport properties and capacitance characteristics of the interface between freshly exfoliated flakes of 1T-TaS$_2$ in intimate van der Waals contact with \textit{n}-type GaAs substrates. The extracted barrier parameters (ideality, barrier height and built-in potential) experience pronounced changes across the Mott-CDW transition in the 1T-TaS$_2$. The CDW-induced changes in barrier properties are well described by a bond polarization model which upon decreasing temperature gives rise to an increased potential drop across the interfacial region due to the localization of carriers and a decreased dielectric constant. 
\end{abstract}
\maketitle

\newpage

\begin{figure}
\includegraphics[width=16cm]{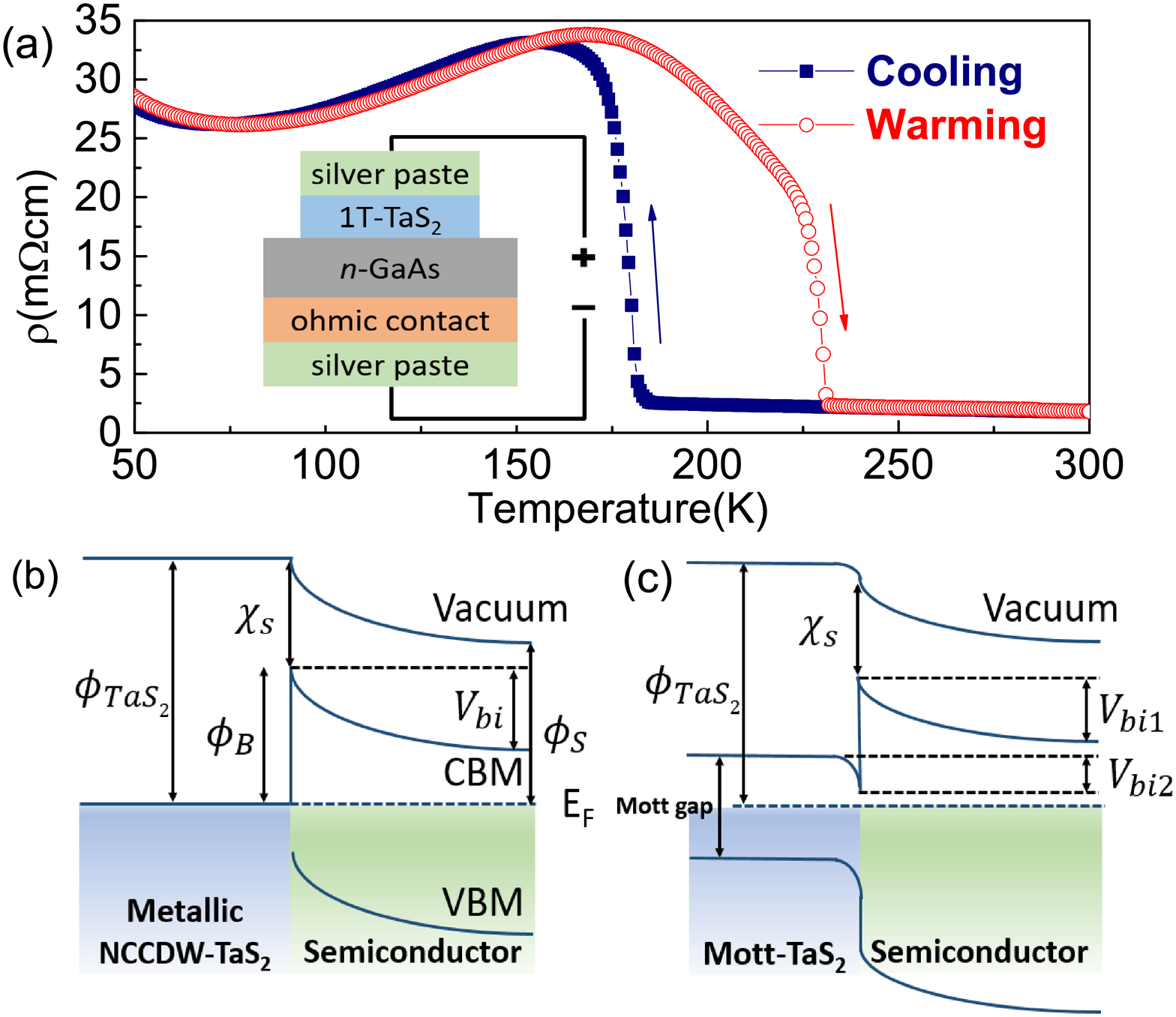}
\caption{(a) Temperature dependence of resistivity for 1T-TaS$_2$ thin flakes. The blue solid squares and red open circles represent the cooling and warming cycle with transition temperatures denoted as T$_{cc}$ and T$_{cw}$ respectively. Inset: Schematic of sample structure with Ag paste contacts. Ideal band alignment for 1T-TaS$_2$/\textit{n}-type semiconductor junctions for 1T-TaS$_2$ in (b) metallic phase (c) Mott-CCDW phase.}
\label{pic1}
\end{figure}

1T-TaS$_{2}$, although simple in structure, embodies various collective electronic states, which have recently triggered intensive study\cite{bishop2008, yu2015, yoshida2015, vask2016}. Above 540 K, 1T-TaS$_{2}$ is a normal metal with the Ta 5d band crossing the Fermi level. Upon cooling, it transforms into an incommensurate CDW (ICCDW)  state, which then evolves into a near commensurate CDW (NCCDW) phase at 350 K. Below T$_{cc}$ ($\sim180$ K), the transition temperature on cooling, a long range ordered commensurate CDW (CCDW) develops accompanied by a large resistance increase. The resistance drops back at T$_{cw}$ ($\sim230$ K), the transition temperature on warming, as shown in Fig.~\ref{pic1}(a). Since 1T-TaS$_2$ is host to a rich variety of CDW phases, there has been widespread interest and success in controlling the transition between CDW states or even hidden metastable states with electric fields, voltage pulses or optical excitations\cite{Holl2015, sto2014, do2015}. These studies not only facilitate the understanding of the correlation between these complex CDW phases in 1T-TaS$_2$, but also imply promise for technological applications. Recently, memristors and oscillators based on 1T-TaS$_2$ utilizing the first-order CDW phase transitions have been realized\cite{yoshida2015, liu2016}. \\
\indent 
Despite the intense effort in studying intrinsic single crystals, 1T-TaS$_2$ based heterojunctions via integration with three-dimensional (3D) semiconductors have not been explored. Our strategy in addressing how 1T-TaS$_2$ might ultimately be used in pure 2D electronic configurations is based on our previous discovery that exfoliated flakes of highly ordered pyrolytic graphite (HOPG) simply pressed against clean semiconductor substrates (Si, GaAs, 4H-SiC) display excellent Schottky barrier rectification effects\cite{tongay_apl}. With the realization that the outermost layer, a single sheet of graphene, in contact with the semiconductor, played a major role in the formation of the Schottky barrier, it was a straightforward matter to produce similar results using 2D single layers of graphene rather than 3D flakes\cite{tongay_prx}. With the current explosion of interest in other van der Waals (vdW) crystals such as layered superconductors, topological insulators and transition metal dichalcogenides, interfacing 2D materials with 3D traditional semiconductors deserves attention and has achieved progress in both high performance devices and detailed device physics prototypes which require modification of the conventional models based on 3D materials\cite{xu2016}. 

In this letter we report on a similar strategy in an investigation of the electrical properties of 1T-TaS$_2$ flakes in intimate contact with \textit{n}-type GaAs substrates. The objective is to obtain a better understanding of the intricate physics at interfaces when long range ordered CDW phases are present in one of the electrodes, in this case the 1T-TaS$_2$. When 1T-TaS$_2$ crystalline flakes make intimate contacts to \textit{n}-type GaAs, strong rectifying current density-voltage ($J-V$) characteristics persist from 100 K to 300 K. Both transport and capacitance measurements are sensitive to the CDW transition (CCDW-NCCDW) during cooling and warming cycles. We interpret our observations within the framework of the bond polarization model where the potential across the interfacial dipole is affected by the presence of CDW phases in 1T-TaS$_2$. Our investigation incorporating charge density waves not only opens up an insightful perspective in understanding the physics at 2D/3D interfaces, but also accelerates the pace of device applications of such systems with first-order phase transitions.  

A schematic diagram of a van der Waals junction is shown in the inset of Fig.~\ref{pic1}(a). Commercially available \textit{n}-type GaAs wafers with nominal $3\sim6\times 10^{16}$ cm$^{-3}$ (Si) doping concentration were used. High quality ohmic contacts were made on the unpolished side with traditional recipes existing in the literature\cite{gaas, sze2006}. 1T-TaS$_2$ thin flakes (10-50 \textmu m thick) were mechanically exfoliated from 1T-TaS$_2$ crystals and then transferred onto cleaned (100) GaAs wafers. For electrical contact, silver paste with an area ranging from 0.3 to 2 mm$^{2}$ covered the top surface of the 1T-TaS$_2$ electrode. We found that samples fabricated in ambient atmosphere or glove boxes containing inert gas showed similar electrical features, implying oxidation effects are not severe during our sample fabrication process (see supplementary material).\\
\indent
All the transport measurements were carried out inside a physical property measurement system (PPMS) using either a Keithley 2400 source meter or lock-in techniques. Hall effect measurement shows the doping concentration of our \textit{n}-type GaAs to be $4.1\times 10^{16}$ cm$^{-3}$ at room temperature. Three-terminal capacitance measurements using an HP 4284A LCR meter were performed on samples in a custom capacitance probe with a sensitivity of 0.1 pF\cite{tongay2009}.

We start with the ideal band alignment of 1T-TaS$_{2}$/\textit{n}-type GaAs junctions. Above T$_{cc}$ during the cooling cycle, 1T-TaS$_{2}$ is in the metallic state with an electron concentration ($\sim$ 10$^{22}$ cm$^{-3}$)\cite{p2006, inda1980}, leading to Schottky like junctions when interfacing with GaAs, as shown in Fig.~\ref{pic1}(b). Transport across such Schottky junctions is dominated by electrons and is well described by the thermionic-emission equation for the current density:
\begin{equation}
J(V,T) =  A^{\ast}T^{2}{\rm{exp}}(-\Phi_{B}(T)/k_{B}T) \times [{\rm{exp}}(e(V-I(V,T)R_s(T))/\eta(T)k_{B}T)-1], \\
\label{thermionic}
\end{equation} 
where $I(V,T)$ is the total current equal to the product $J(V,T)$ with the effective area $A$,  $A^{\ast}$ is the Richardson constant, $T$ is the temperature, $\Phi_{B}(T)$ is the zero bias Schottky barrier height, $\eta(T)$ is the ideality factor indicating the contribution of thermionic-emission processes, and $R_{s}(T)$ is the contact resistance. Complementary to transport, junction capacitance in the reverse bias region is given by
$ A^2/C^2=2(V_{bi}(T) - V)/eN_D \epsilon_{semi}\epsilon_{0}$,
where $A$ is the effective sample area, $V_{bi}$ is the built-in potential, $N_D$ is the ionized donor density, $\epsilon_{semi}$ is the relative dielectric constant of the semiconductor and $\epsilon_{0}$ is the vacuum permittivity.
Compared to $J-V$ measurements, capacitance offers a more accurate band profile, but is insensitive to various transport channels.\\
\indent
When entering the Mott-CCDW state, 1T-TaS$_2$ becomes a \textit{p}-type semiconductor, with a Mott gap $\sim300$ meV opening\cite{inda1980, cho2015}. Thus, 1T-TaS$_2$ in the Mott-CCDW state will form pn heterojunctions with \textit{n}-type GaAs. The corresponding band alignment is demonstrated in Fig.~\ref{pic1}(c)\cite{you2013}. As for transport, tunneling current is affected as the density of states (DOS) profile changes on the 1T-TaS$_2$ side. Holes also contribute to transport. 

Junction capacitance still leads to an accurate measure of the built-in potential, which is given by\cite{sze2006}:
\begin{equation}
\dfrac{A^2}{C^2}=\dfrac{2(V_{bi}(T) - V)}{e}(\dfrac{1}{N_D \epsilon_{semi}\epsilon_{0}}+\dfrac{1}{N_A\epsilon_{TaS_2}\epsilon_{0}}), 
\label{heterojunction capacitance}
\end{equation}
where $N_A$ is the effective acceptor density in 1T-TaS$_2$. In contrast to Schottky barriers, the measured $V_{bi}$ equals the sum of band bending on the 1T-TaS$_2$ side and GaAs side, denoted by $V_{bi1}$ and $V_{bi2}$ respectively in Fig.1(c). Given that 
$V_{bi1}/V_{bi2}=N_A\epsilon_{TaS_2}/N_D \epsilon_{semi}$\cite{sze2006},
the band bending in 1T-TaS$_2$ is still very small, since the carrier concentration of 1T-TaS$_2$ in the Mott-CCDW state ($\sim$ 10$^{19}$ cm$^{-3}$)\cite{inda1980} is much larger than that of GaAs ($\sim$ 10$^{16}$ cm$^{-3}$) and the dielectric constant of both sides is similar\cite{bishop2008}.

\begin{figure}
\includegraphics[width=16cm]{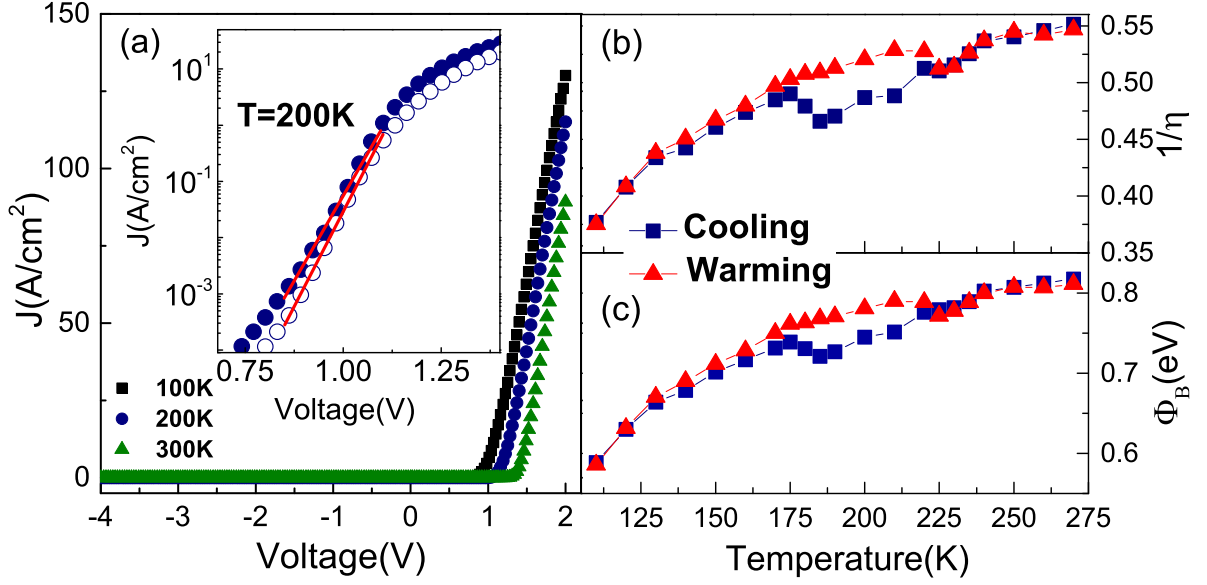}
\caption{(a) The $J - V$ characteristics of 1T-TaS$_2$/GaAs junctions measured at 100 K, 200 K and 300 K in the cooling cycle. Inset: Semi-logarithmic plots of the forward bias $J - V$ curves at 200 K in the cooling (blue solid circles) and warming (blue open circles) cycle. The red solid lines are used to fit with the thermionic-emission equation. Temperature dependence of (b) 1/$\eta$ and (c) zero-bias barrier height $\Phi_B$ are shown for both the cooling (blue squares) and warming (red triangles) cycles.}
\label{APLIV}
\end{figure}

Fig.~\ref{APLIV}(a) shows the typical $J-V$ curves of a 1T-TaS$_2$/GaAs junction below and above the transition temperature in the cooling cycle. 1T-TaS$_2$ is defined as the positive side. All $J-V$ curves show a pronounced rectifying feature. The reduction of forward bias current density at low temperature is mainly due to the increasing contact resistance. To explore how CDW formation would affect the transport properties, we plot in the inset of Fig.2(a) the forward bias $J-V$ curves at 200 K, where almost the largest hysteresis of 1T-TaS$_2$ single crystal resistivity takes place. A reduction of forward bias current is observed in the warming cycle.\\
\indent
In principle, transport data of 1T-TaS$_2$/GaAs junctions cannot be fitted with a unified model across CDW phase transitions. However, the thermionic-emission process, although not the sole contribution, still plays an important role. Thus, analysis of $J-V$ curves based on the thermionic-emission equation  
still suffices to reveal changes at the interface. The fitted $\Phi_B$ is thus a physical parameter incorporating the most dominant transport mechanisms\cite{ang2016}. As shown by the red dashed line in the inset of Fig.~\ref{APLIV}(a), typical $J-V$ data in the forward bias region displays over 2 decades of linearity on a semilogarithmic scale, enabling us to extract $\eta (T)$ and $\Phi_B(T)$.

As shown in Fig.~\ref{APLIV}(b), the ideality of 1T-TaS$_2$/GaAs junctions generally increases with decreasing temperatures primarily due to the significant contribution of tunneling and thermally assisted tunneling at low temperatures\cite{pa1965}. When going through the NCCDW to Mott-CCDW transition, $\eta$ experiences a rapid drop (Fig.~\ref{APLIV}(b)), while, $\Phi_B$ on the other hand, increases at T$_{cc}$ (Fig.~\ref{APLIV}(c)). Our explanation is that with lowering temperature the Mott gap opens, thereby lowering the available density of states near the Fermi level and thus the tunneling current. Consequently, the contribution of thermionic process is more dominant, represented by a drop in $\eta$ concomitant with an increase of $\Phi_B$, both signifying an increased contribution of thermionic emission current\cite{we1991}. Although transport channels like hole diffusion and generation-recombination may exist with Mott gap opening, according to our data, their contribution is less significant than tunneling processes. 
The obvious hysteresis in both $\eta$ and $\Phi_B$ demonstrates that the formation of the Mott-CCDW state indeed modifies the transport picture of 1T-TaS$_2$/GaAs junctions.

\begin{figure}
\includegraphics[width=16cm]{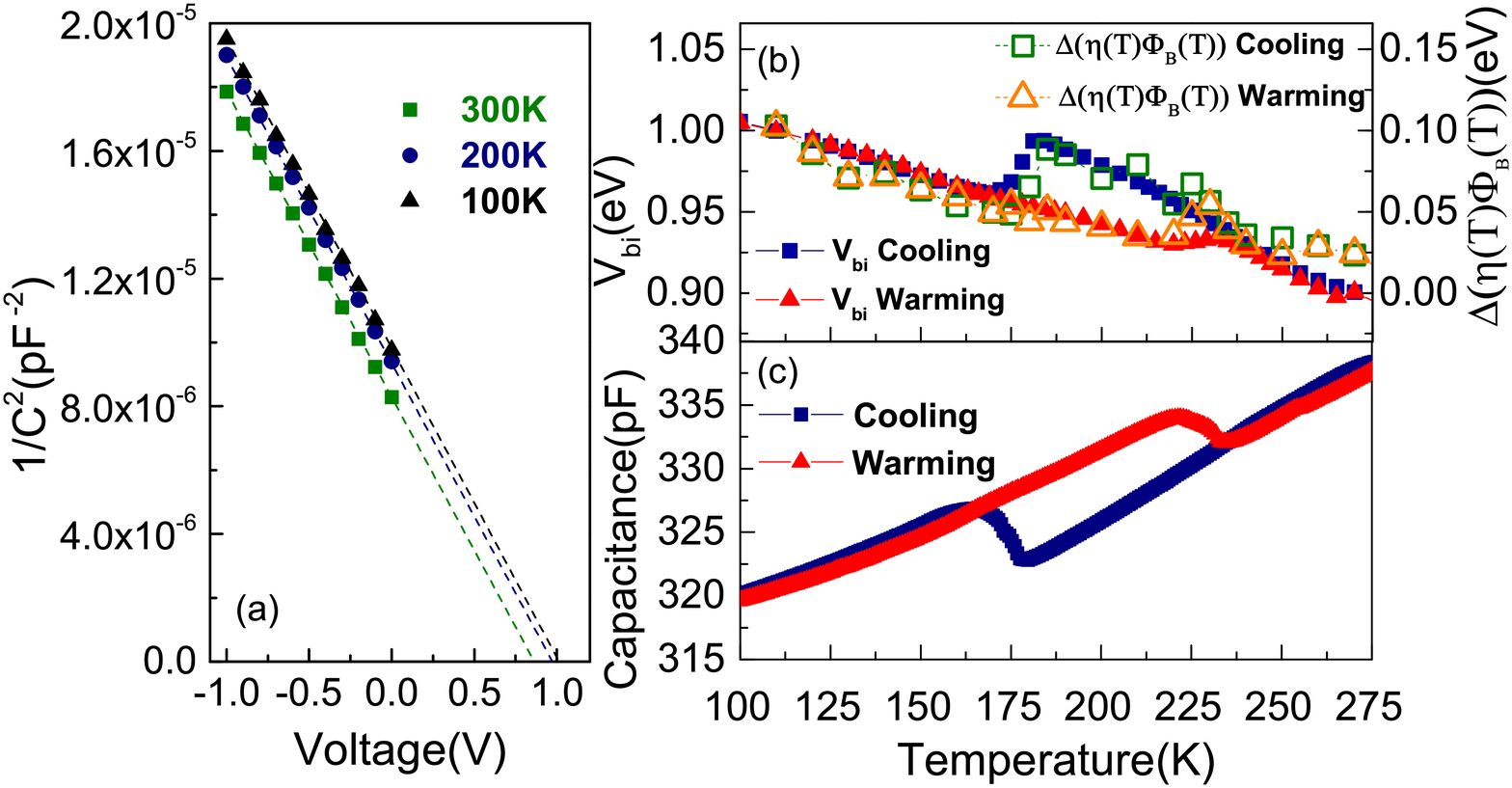}
\caption{(a) $1/C^2 - V$ plots of the 1T-TaS$_2$/GaAs junction measured at 100 K, 200 K and 300 K in the cooling cycle. (b) Temperature evolution of the built-in potential V$_{bi}$ in the cooling (blue squares) and warming (red triangles) cycle. Change of flat band voltage as a function of temperature in the cooling (olive open squares) and warming (orange open triangles) cycle. (c) Temperature dependent zero bias capacitance for the 1T-TaS$_2$/GaAs junction in the cooling (blue squares) and warming (red triangles) cycle.}
\label{APLCVCT}
\end{figure}

Capacitance reflects the flat band voltage\cite{we1991}, giving a more accurate description of interface band alignment evolution across the CDW phase transition. The characteristic linear relationship between $1/C^2$ and $V$ is observed in our studied temperature range as shown in Fig.~\ref{APLCVCT}(a). Extrapolation of these data to the abscissa provides temperature-dependent $V_{bi}(T)$ as summarized in Fig.~\ref{APLCVCT}(b). For normal metal/\textit{n}-type semiconductor junctions, as temperature decreases, V$_{bi}$ keeps increasing, which is consistent with the work function of \textit{n}-type extrinsic semiconductors which decreases when lowering the temperature. Our extracted V$_{bi}$ values exhibit the same trend except the hysteresis and rapid change at the CDW transition temperatures. The measured V$_{bi}$ drops by around 50 meV during cooling near 180 K and jumps back on warming near 230 K. 
\indent

As previously reported, the product $\eta(T)\Phi_B(T)$ is an estimation of flat band voltage\cite{ang2016, we1991}. We also plot as open symbols the $T$-dependent product difference, $\Delta\eta(T)\Phi_B(T)=\eta(T)\Phi_B(T)-\eta(300K)\Phi_B(300K)$, in Fig.~\ref{APLCVCT}(b) and find the same temperature dependence as V$_{bi}(T)$ (solid symbols) obtained from capacitance measurements. This similarity suggests our transport and capacitance data are consistent with each other as shown previously for a variety of diodes\cite{ang2016, we1991}. The abrupt decrease of $V_{bi}$ is further confirmed by zero-bias capacitance temperature ($C-T$) measurement as shown Fig.~\ref{APLCVCT}(c). When entering the Mott-CCDW state, if no built-in potential (effective work function) variation takes place, the junction capacitance would slightly decrease as the band bending in 1T-TaS$_2$ side results in two capacitors connected in series. However, the measured capacitance changes in the opposite way (i.e., increases), implying a decrease in $V_{bi}(T)$, as predicted by Eq.~\ref{heterojunction capacitance}. A similar picture applies when 1T-TaS$_2$ changes back to the metallic state, with a jump of $V_{bi}(T)$ as long range CDW order is destroyed. We note that junctions fabricated in the glove box demonstrate similar hysteresis features in both transport parameters and capacitance data (see supplementary material).

To explain our observed capacitance features, a detailed interfacial band alignment is needed. Considering the ideal metal-semiconductor (MS) interface, the Shottky-Mott rule applies, which predicts the room temperature barrier height to be 1.13 eV, given that the work function of 1T-TaS$_2$ is 5.2eV\cite{shimada1994} and the electron affinity of GaAs is 4.07 eV\cite{sze2006}. However, based on our built-in potential, the average barrier height of 8 junctions at room temperature is 0.97 eV, indicating the necessity of considering Fermi level pinning (FLP) effect commonly observed in real MS interfaces\cite{tung2001}. The most relevant mechanisms explaining FLP effect include metal-induced gap states (MIGS)\cite{metal-Si, h1965}, defect induced gap states (DIGS)\cite{h1983} and bond polarization\cite{tung2000, tung20012}. However, MIGS has been proven to be largely suppressed in van der Waals junctions\cite{sc2016}, while DIGS is closely correlated to the quality of 1T-TaS$_2$ flakes and semiconductor wafers. We assume that the DIGS effect, although present, is small and cannot serve as the explanation for our observed features taking place right at T$_{cc}$ and T$_{cw}$. 

\begin{figure}
\includegraphics[width=16cm]{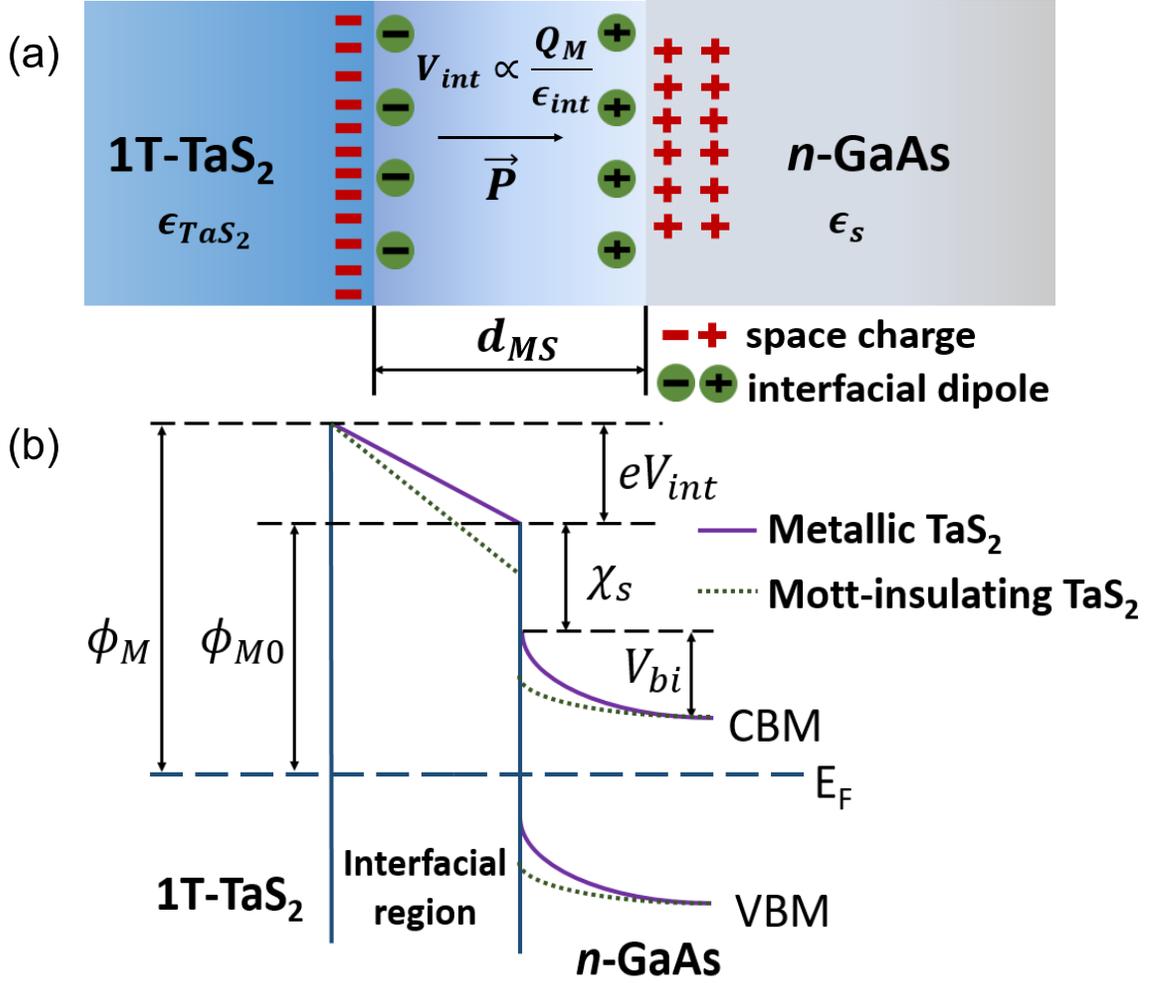}
\caption{(a) The cross-sectional view of the 1T-TaS$_2$/GaAs interface based on bond polarization model. (b) The band alignment picture 
for the metallic 1T-TaS$_2$/GaAs interface (purple solid lines) and Mott-insulating 1T-TaS$_2$/GaAs interface (green dotted lines). }
\label{pic4}
\end{figure}

\indent
In the bond polarization model, as shown in Fig.~\ref{pic4}(a), in addition to the formation of the space charge region, charge exchange also occurs between the topmost layer of metal and semiconductor when an intimate contact is formed. DFT calculations confirm such charge transfer in transition metal dichalcogenide based interfaces\cite{sc2016, g2014}. 

According to the bond polarization model, charge transfer between 1T-TaS$_2$ and the semiconductor can be written as:
\begin{equation}
eQ_{M}=e(\Phi_{M}-\chi- E_{g}/2)/(E_{g}+\kappa), 
\label{BP}
\end{equation} 
where $E_{g}$ is the band gap of semiconductor, $\Phi_{M}$ is the work function of intrinsic 1T-TaS$_2$, $\chi_{s}$ is the electron affinity of the semiconductor and $\kappa$ is a parameter related to interfacial Coulomb interaction\cite{tung2014}. Since 1T-TaS$_{2}$ has a large work function, the interfacial charge transfer would result in an additional dipole pointing from 1T-TaS$_2$ to the semiconductor. Such an interfacial dipole, attached to 1T-TaS$_2$, would modify the work function of 1T-TaS$_2$ by $eV_{int}$\cite{l2003}, where $V_{int}=-eN_{B}d_{MS}Q_{M}/\epsilon_{int}$, with $e$ the electric charge, $\epsilon_{int}$ the effective interface dielectric constant and $eN_{B}d_{MS}Q_{M}$ representing the interfacial dipole strength with areal density $N_{B}$ and separation $d_{MS}$ ($\sim$3 {\AA})\cite{g2014}. Here we define the effective work function ($\Phi_{M0}$) of 1T-TaS$_2$, which includes the interfacial region, as $\Phi_{M0}=\Phi_{M}+eV_{int}$.\\

\indent
The measured built-in potential for both MS junctions and pn heterojunctions is given by the difference between work functions. One apparent explanation for our observed built-in potential drop in the cooling cycle takes into account the work function change of intrinsic 1T-TaS$_2$, since \textit{n}-type GaAs experiences no sudden change of the work function across the CDW phase transition. However, limited information about the 1T-TaS$_{2}$ work function is available. Only Shimada claims that no noticeable change of 1T-TaS$_2$ work function is observed across the CDW phase transition\cite{shimada1994, shimadad}. As is well known, the extraction of work function is always intricate and can even provide opposite results across phase transitions\cite{you2013}, which makes the intrinsic work function change argument less convincing. 

Another more promising explanation addresses the change of interfacial potential $eV_{int}$ across CDW phase transitions. Since the atomic displacement for 1T-TaS$_2$ changes less than 7$\%$\cite{ro2011}, we do not expect a sudden change in the number of interfacial dipoles relying heavily on lattice mismatch. The only remaining variable $\epsilon_{int}$, determining the potential drop across a given interfacial dipole strength, is a combination of dielectric constant of GaAs and 1T-TaS$_2$. The dielectric constant of GaAs remains almost unchanged in the studied temperature range. Above T$_{cc}$ during the cooling cycle, metallic 1T-TaS$_2$ has a large dielectric constant. When entering the Mott-CCDW state, the dielectric constant of 1T-TaS$_2$ becomes comparable to GaAs, consistent with a metal-insulator transition\cite{bishop2008}. Such a dielectric constant drop leads to a larger interfacial  potential drop as shown in Fig.~\ref{pic4}(b), decreasing the effective work function of 1T-TaS$_2$. Accordingly, the measured V$_{bi}$ experiences a drop at T$_{cc}$ and then jumps back at T$_{cw}$.\\

\indent
In summary, we have systematically studied the transport properties and capacitance characteristics of 1T-TaS$_{2}$/\textit{n}-type GaAs van der Waals heterojunction. The transport related parameters $\eta$ and $\Phi_{B}$ are sensitive to the Mott-CCDW formation. We ascribe the transport picture modification to different tunneling contributions originating from a modified DOS profile. In addition, the built-in potential from $C-V$ measurement drops with the CCDW formation which is consistent with the concomitant changes in the zero-bias $C-T$ data. We explain the observed feature based on bond polarization model that localization of free carriers renders a larger potential drop across interfacial dipoles due to less effective screening. Such combination of exfoliated TMD materials with traditional semiconductors may bridge the gap between traditional MS interfaces and pure 2D material interfaces where large scale fabrication and ohmic contact remain as challenges and have implication for the next generation devices based on phase transition TMDs.\\

\indent
See supplementary material for the electrical measurements of a typical junction fabricated in the glove box containing argon.\\

\indent
The authors thank Haoming Jin, Jie Hou and Chris Samouce for assistance with sample preparation and the Nanoscale Research Facility at the University of Florida for access to rapid thermal annealing facilities. This work was supported by the National Science Foundation under the Division of Materials Research Grant No. DMR-1305783 (AFH) and the Department of Energy under Grant No. DE-FG02-86ER45268 (GRS).

\newpage

\clearpage

\section*{S\MakeLowercase{upplemental information for} D\MakeLowercase{etection of charge density wave phase transitions at} 1T-T\MakeLowercase{a}\MakeLowercase{S$_2$}/G\MakeLowercase{a}A\MakeLowercase{s interfaces}}

The electrical measurement data are shown for one typical 1T-TaS$_2$/GaAs junction fabricated in the glove box containing argon. Transport parameters (Figure S 1) and capacitance data (Figure S 2) are consistent with the data of the sample fabricated in the ambient atmosphere which are presented in the main text. The clear hysteretic features in barrier parameters suggest that the CDW induced changes at the interface are robust and the surface degradation effect of 1T-TaS$_2$ is small during our sample fabrication process.

\setcounter{figure}{0}
\renewcommand{\figurename}{FIG. S}
\clearpage
\begin{figure}
\includegraphics[width=16cm]{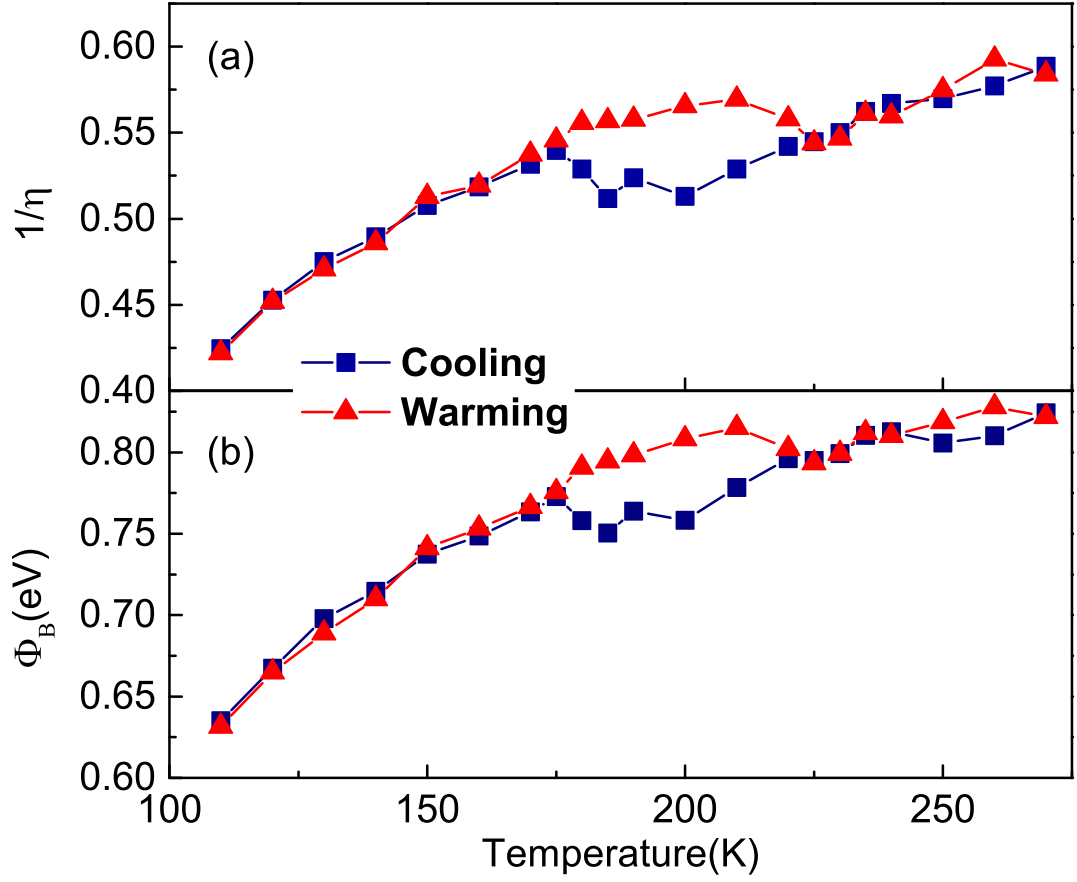}
\caption{Temperature evolution of (a) 1/$\eta$ and (b) $\Phi_B$ extracted from the thermionic-emission equation fitting for both the cooling (blue squares) and warming (red triangles) cycles. }
\label{pics1}
\end{figure}

\begin{figure}
\includegraphics[width=16cm]{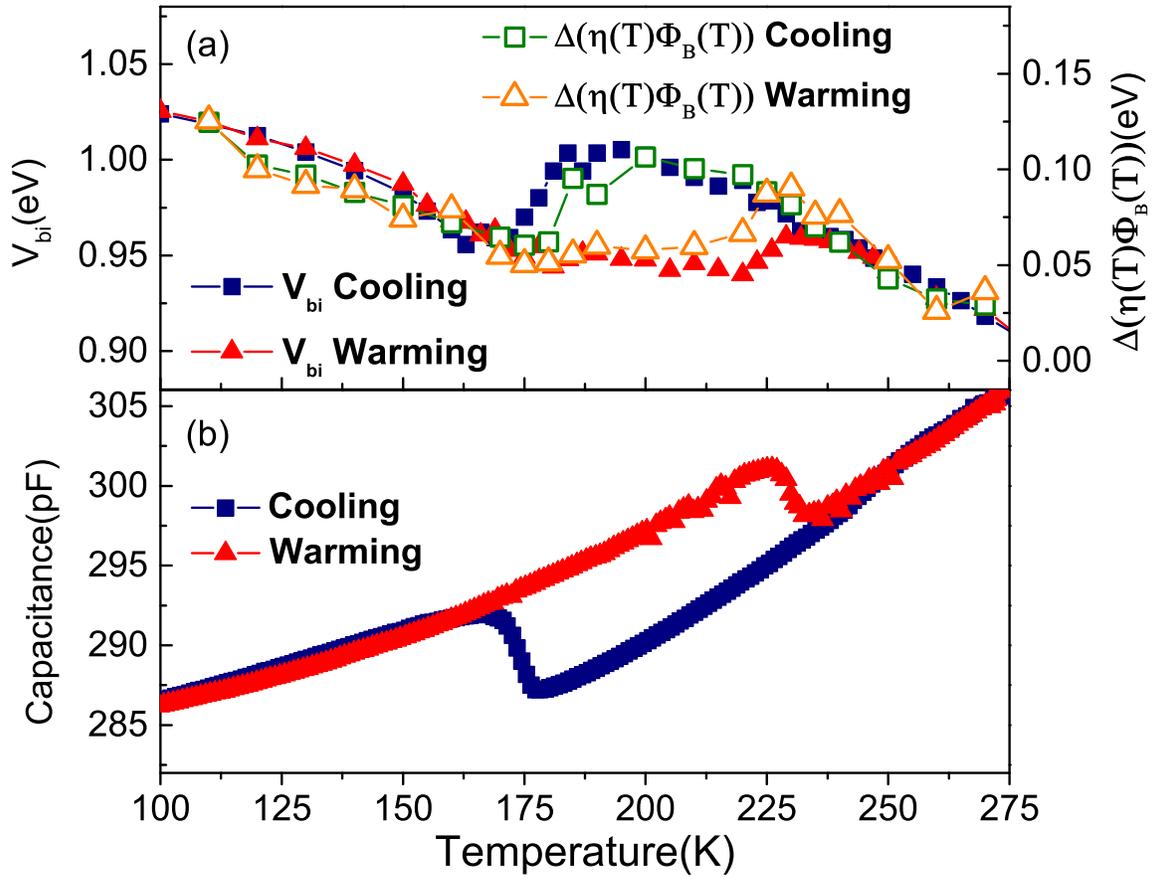}
\caption{(a) Temperature dependence of V$_{bi}$ in the cooling (blue squares) and warming (red triangles) cycles. Variation of ﬂat band voltage as a function of temperature in the cooling (olive open squares) and warming (orange open triangles) cycles. (b) Temperature dependent zero bias capacitance in the cooling (blue squares) and warming (red triangles) cycles.}
\label{pics2}
\end{figure}
\end{document}